\documentclass{PoS}
\newcommand{\DBD}{0$\nu$DBD}

\newcommand{\TEO}{$\mathrm{TeO}_2$}
\newcommand{\TEHT}{$^{130}\mathrm{Te}$}

\newcommand{\Cuoricino}{CUORICINO}
\newcommand{\Cuore}{CUORE}

\providecommand*{\un}[1]{\ensuremath{\mathrm{~#1}}}
\usepackage{overpic}
\usepackage{comment}

\title{Feasibility study of dark matter searches with the CUORE experiment}

\ShortTitle{Dark Matter searches with CUORE}

\author{\speaker{M.~Vignati}, on behalf of the CUORE collaboration\\
        Sapienza Universit\`a di Roma and INFN Sezione di Roma, Roma, I-00185, Italy\\
        E-mail: \email{marco.vignati@roma1.infn.it}}

\abstract
{CUORE will be a 1 ton experiment made of about 1000 \TEO\ bolometers.
It will probe the neutrinoless double beta decay (\DBD) of \TEHT,
a tool to test the neutrino nature and mass.
The excellent energy resolution and the low background of these
detectors will make CUORE a leading experiment in this field,
improving the sensitivity to the half-life of \DBD\ by more than an order of magnitude.
Bolometric detectors, however, are also sensitive to nuclear
recoils and can be used to search for dark matter interactions. 
In principle CUORE, thanks to its mass, could look for an annual modulation 
of the counting rate at low energies.
We developed a trigger and a pulse shape identification algorithm,  
that allow to lower the energy threshold down to the few keV region. 
We present the preliminary results obtained on an array made of 
four CUORE-like crystals, and the prospects for a dark matter search in CUORE.
}

\FullConference{Identification of Dark Matter 2010-IDM2010\\
		July 26-30, 2010\\
		Montpellier France}

\begin{document}
\section{The CUORE experiment}
\begin{figure}[b]
\centering
\includegraphics[width=0.55\textwidth]{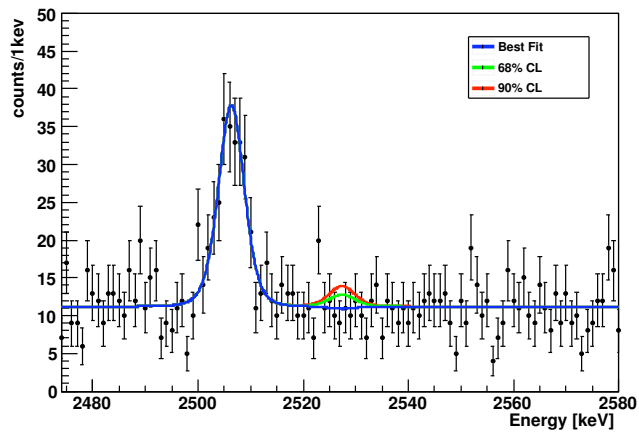}
\label{fig:qino}
\caption{\Cuoricino\ energy spectrum in the region of interest. The line at 2505\un{keV} is due to $^{60}$Co contamination of the materials
surrounding the bolometers. No peak is found at the expected \DBD\ energy of 2527\un{keV}.}
\end{figure}

The \Cuore\ experiment is designed to search for the  neutrinoless double beta decay (\DBD)
of \TEHT~\cite{Ardito:2005ar}. The observation of this nuclear decay would determine that
the neutrino is a Majorana particle, unlike all other fermions that are Dirac particles, and would set
a scale for the absolute value of the neutrino mass.  
\Cuore\  will be made of 988 \TEO\ bolometers of 750\un{g} each. 
Bolometers are detectors in which the energy from particle interactions is converted
into heat and measured via the resulting rise in temperature.
Operated at a temperature of about 10\un{mK}, these
detectors maintain an energy resolution of a few keV over their energy
range, extending from a few keV up to several MeV.  The measured resolution
in the region of interest (2527\un{keV}) is about 5\un{keV\,FWHM}; this,
together with the low background and the high mass of the experiment,
determines the sensitivity to the \DBD. 
\Cuore\ will be installed in the Gran Sasso National Laboratory (LNGS) in Italy
and will start the data taking in about 2 years. The first CUORE tower, CUORE-0, composed by 52 bolometers, is under preparation
and will start the data taking in the next year. 
A demonstrator experiment, \Cuoricino, was operated in the same laboratory in the years 2003-2008.
It was composed by  62 \TEO\ bolometers, for a total mass of 40.7\un{Kg}. The acquired
statistics was 19.75$\,\rm kg (^{130}Te)\cdot y$, and the final result is presented in Fig.~\ref{fig:qino}. 
No \DBD\ signal was found, the
background level was $0.16\pm0.01$ counts/keV/kg/y and the
corresponding lower limit on the \DBD\ half-life of \TEHT\ is
2.8$\times$ 10$^{24}$ y (90\% C.L.). This limit leads to an upper limit
on the neutrino effective Majorana mass ranging from 0.3
to 0.7 eV, depending on the nuclear matrix elements considered in
the computation.

 \Cuore\ could also search for dark matter (DM)
interactions, provided that the energy threshold is sufficiently low. DM
candidates such as weak interacting massive particles (WIMPs) and axion-like particles (ALPs)~\cite{Bertone:2004pz} are expected to
produce signals at energies below $\sim 30\un{keV}$, and the interaction
rate increases as the energy decreases.
The energy threshold of \Cuoricino\ bolometers, achieved using a trigger algorithm
applied to the raw data samples, was of the order of tens of keV. 
If the threshold were of few keV, the \Cuore\ experiment could be sensitive to DM interactions
and thus play an important role in this growing research area. In the following we present a method we developed
to lower the energy threshold, and the projected \Cuore\ sensitivity to WIMP spin independent interactions.

\section{Experimental setup}

A \Cuore\ bolometer is composed of two main parts,
a \TEO\ crystal and a neutron transmutation doped Germanium (NTD-Ge) thermistor~\cite{Itoh:1996}. The crystal is
cube-shaped (5x5x5\un{cm^3}) and held by Teflon supports in copper frames.  The frames
are connected to the mixing chamber of a dilution refrigerator, which
keeps the system at the temperature of $\sim 10\un{mK}$.  The thermistor
is glued to the crystal and acts as thermometer. When energy is released
in the crystal, its temperature increases and changes the thermistor's
resistance.  The thermistor is
biased with a constant current, and the voltage across it constitutes the
signal~\cite{AProgFE}.  Usually, a Joule heater is also glued to the crystal.
It is used to inject controlled amounts of energy into the crystal, to emulate
signals produced by particles~\cite{Arnaboldi:2003yp}.
The signal is amplified, filtered with a 6-pole active Bessel filter with
a cut-off frequency ranging between 12 and 20\un{Hz}, and then acquired with an 18-bit ADC
with a sampling frequency of 125\un{Hz}.

\begin{figure}[hb]
\centering
\includegraphics[width=0.5\textwidth]{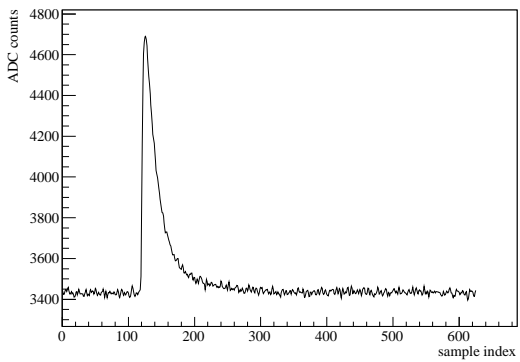}
\caption{Signal produced by an 88\un{keV} $\gamma$
particle fully absorbed in one detector.  The signal is sampled at
125\un{Hz} and the length of the window corresponds to 5.008\un{s}.
\label{fig:signal_88} 
}
\end{figure}

The data presented in this paper comes from a four-bolometers array operated by the \Cuore\ collaboration at LNGS in 2009. 
The main purpose of the detector was to check one of the first 
production batches of \Cuore\ crystals~\cite{ioanprod}.  Thermistors were glued on all crystals,
while heaters were glued only on two crystals.
A typical signal produced by an 88\un{keV} $\gamma$ particle
generated by a $^{127m}$Te de-excitation in a crystal is shown in
Fig.~\ref{fig:signal_88}. 

\section{Lowering the energy threshold}
 At low energies the detection capability is limited by the detector noise.
Electronics spikes, mechanical vibrations, and temperature fluctuations can
produce pulses that, if not properly identified, generate nonphysical
background. The lower the energy released in the bolometer, the
more difficult it is to discriminate between physical and nonphysical pulses.

\begin{figure}[hbtp]
\centering
% run 900216 ch 3 ev 218
\begin{minipage}{0.49\textwidth}
\begin{overpic}[width=1\textwidth]{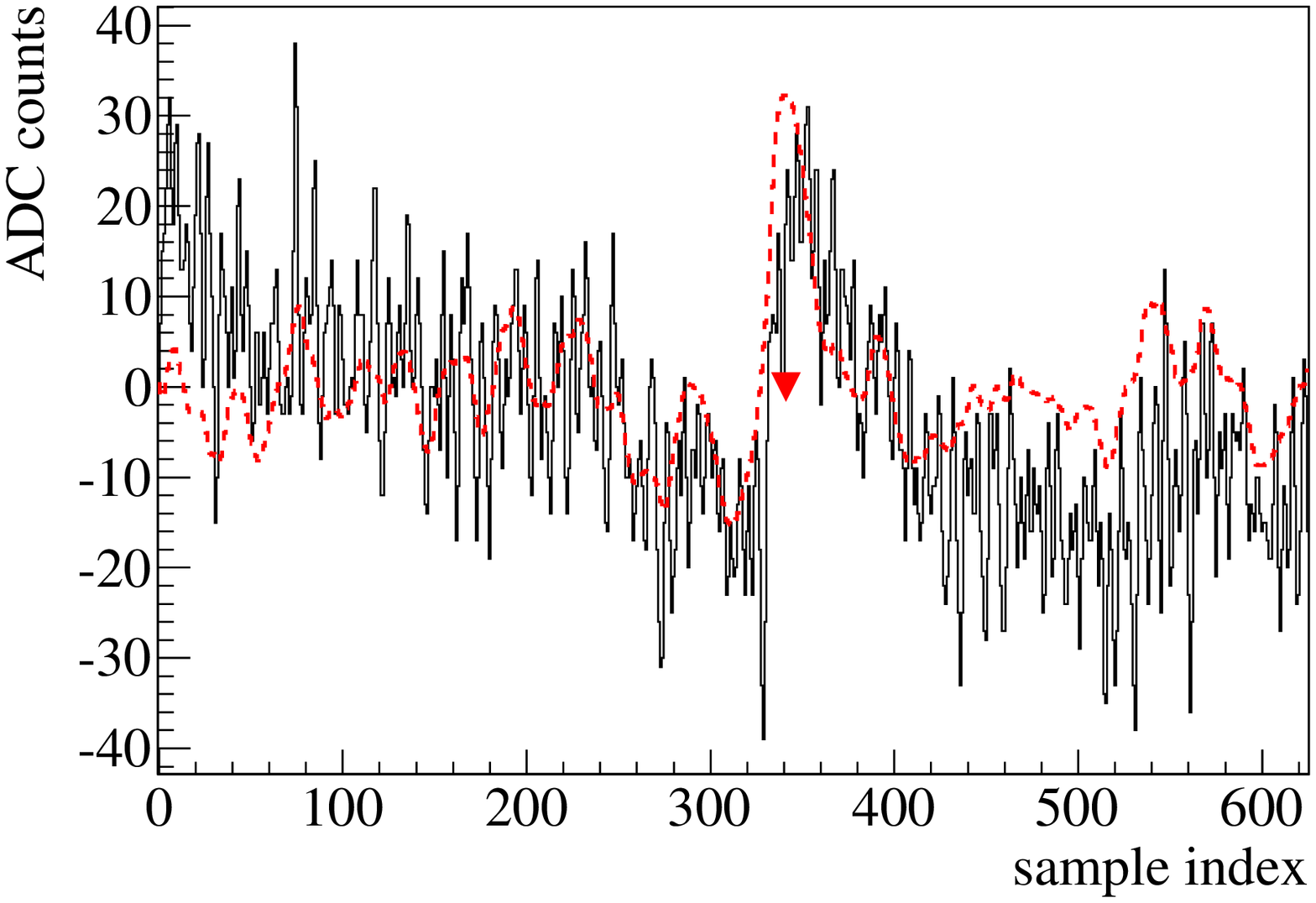}
\put(60,60){\footnotesize 3\un{keV} signal}
\end{overpic}
\end{minipage}
% run 900216 ch 3 ev 830
\begin{minipage}{0.49\textwidth}
\begin{overpic}[width=1\textwidth]{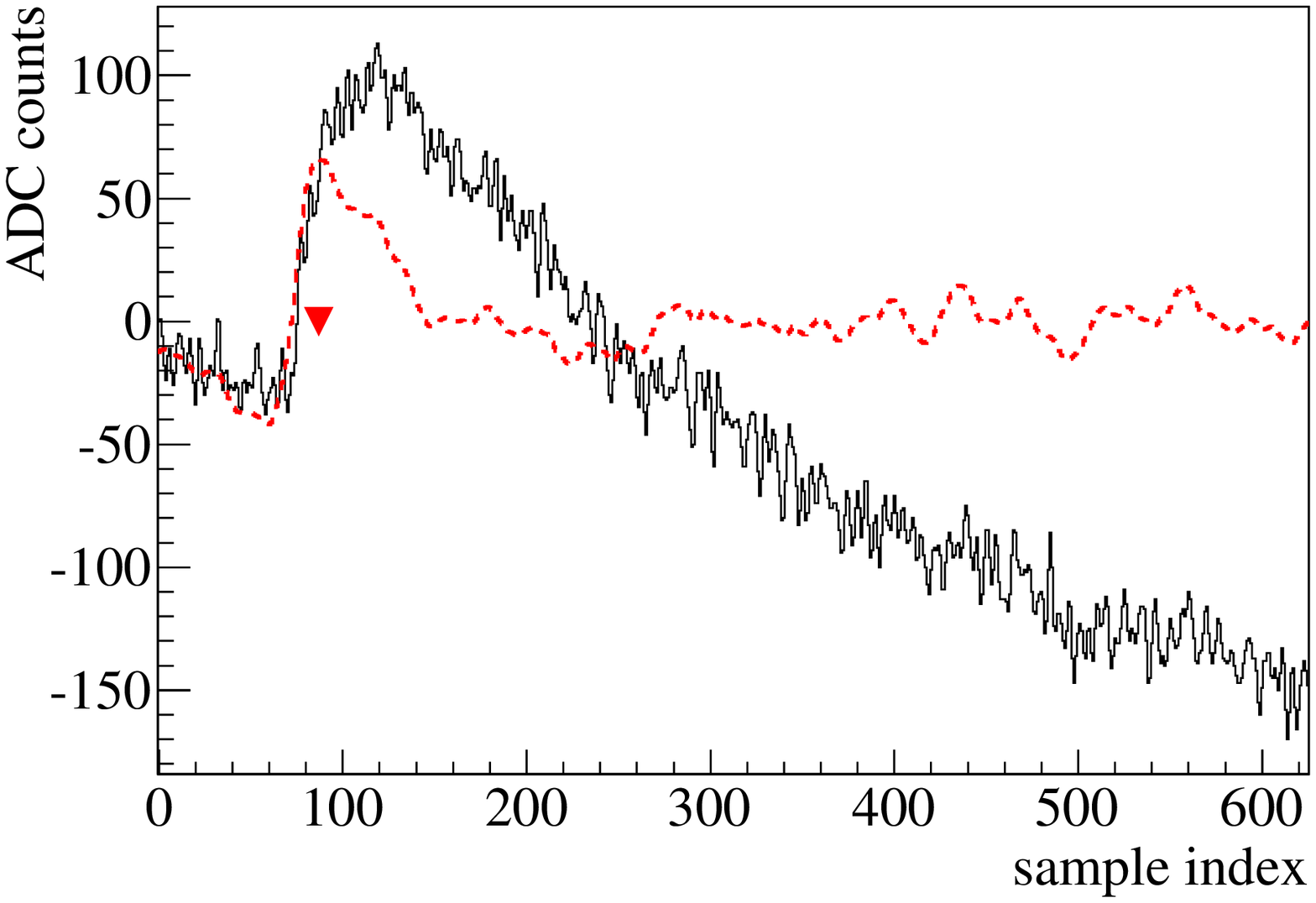}
%\begin{overpic}[width=1\textwidth]{histo_SI_ch3_2.5.eps}
\put(40,60){\footnotesize detector vibration}
\end{overpic}
\end{minipage}
\caption{Windows of data samples containing pulses. The raw data (solid black line) have been shifted such that the first sample has zero ADC counts; the filtered data (dashed red line) are not modified.
Red triangles identify the pulses detected by the trigger algorithm.
The filter removes the baseline drifts and suppresses pulses with different shapes than the expected signal.
}
\label{fig:ot_filt_windows}
\end{figure}

We developed a trigger and  a pulse shape identification algorithm~\cite{otarticle} that
operate on data samples processed using the matched filter
technique~\cite{Gatti:1986cw}. This filter is designed to estimate the amplitude of a signal,
maximizing the signal to noise ratio.
\begin{comment}
The transfer function is matched
to the signal shape, and pulses with different shape are suppressed.
The noise power spectrum of the detector, $N(\omega_k)$, and the signal
shape, $s_i$, are needed to build the transfer function:
\begin{equation}
H(\omega_k) =  h \frac{s^*(\omega_k)}{N(\omega_k)}e^{-\jmath\, \omega_k i_M }\,,
\label{eq:of}
\end{equation}
where $s(\omega_k)$ is the Discrete Fourier Transform (DFT) of $s_i$, $i_M$ is the maximum position of $s_i$
in the acquisition window, and $h$ is a normalization constant that 
leaves unmodified the amplitude of the signal at the filter output. The estimation of $N(\omega_k)$ is made by averaging 
the power spectrum of a large set of data windows acquired with a random trigger, 
while the signal shape $s_i$ is obtained by averaging a set of high-energy particle signals in the 1-3\un{MeV} range.
\end{comment}
Some slices of the filtered data are shown in Fig.~\ref{fig:ot_filt_windows}. In it  one can see that the noise fluctuations
are reduced and  signals with shape different from the expected are suppressed.
\subsection{Detection efficiency}
\begin{figure}[htb]
\centering
\includegraphics[width=0.6\textwidth]{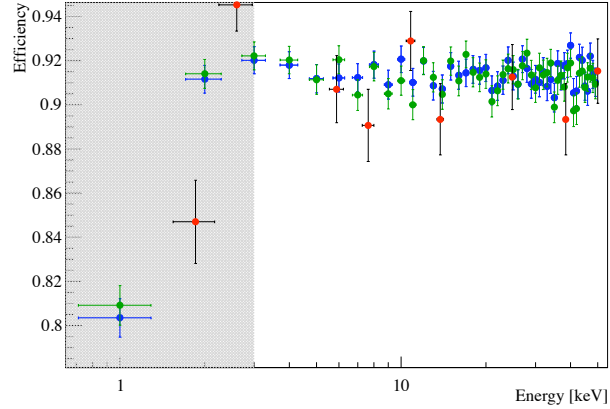}
\caption{Detection efficiency as a function of energy for one of the four bolometers. The red dots represent the detection efficiencies on pulses generated by the heater, the blue (green) dots are simulated heater (particle) pulses. The plateau is reached at $\sim 3\un{keV}$, value which
determines the energy threshold used in the data analysis. Above this threshold simulated data agrees well with real data, validating
the efficiency estimation on bolometers without the heater, where only the simulation can be used.}
\label{fig:efficiency}
\end{figure}

Pulses are triggered when the data samples exceed a positive
threshold, and a new trigger is possible when the data samples return
below threshold, or after that a local maximum is found. The detection efficiency is
estimated on pulses generated by the heater, performing an energy scan from about $1\un{keV}$
up to 100\un{keV}. In the case of bolometers without the heater, the estimation is made
on simulated data. In Fig.~\ref{fig:efficiency} the efficiency as a function of energy is shown for a bolometer with the heater,
and is compared with simulations of heater and particle pulses. In it,  one can see that the simulation agrees well
with data, validating the efficiency estimation on bolometers without the heater. The efficiency reaches 
a plateau of $\sim 91\%$  at $E\sim3\un{keV}$, which is the software energy threshold we set in the data analysis.
Three bolometers has a $3\un{keV}$ energy threshold and a detection efficiency in excess of $80\%$, while the fourth has a threshold of $\sim 10\un{keV}$. To ease the analysis job we ignored the fourth bolometer.

\subsection{Pulse shape identification}
\begin{figure}[htb]
\centering
\includegraphics[clip=true, width=0.55\textwidth]{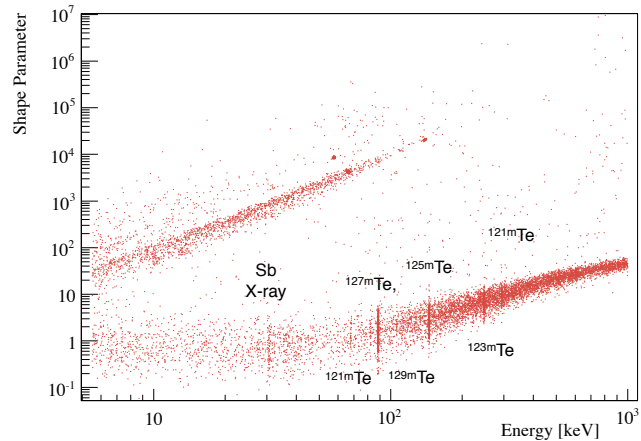}
\caption{Shape parameter distribution of all the events in the range 5-1000\un{keV}
for one of the three bolometers used in the data analysis.
The band at low values of the shape parameter is populated by signal events, where the lines generated by metastable Tellurium
isotopes are visible. At higher values there are triggered mechanical vibrations and spikes.}
\label{fig:shapeparameter}
\end{figure}

To remove nonphysical pulses we fit the filtered pulses with the expected shape of the filtered signal.
The $\rm \chi^2/ndf$ of the fit is used as shape parameter, and the distribution obtained for one of the bolometers is shown in 
Fig.~\ref{fig:shapeparameter}.  In it, one can see that nonphysical pulses are very well separated. We apply a cut on this variable
to select signal events in the energy region below $30\un{keV}$. The probability of loosing a signal event after this cut has
been estimated to be less than $3\%$.

\section{Sensitivity to dark matter interactions}
\begin{figure}[bt]
\centering
\includegraphics[angle=0,width=0.66\textwidth]{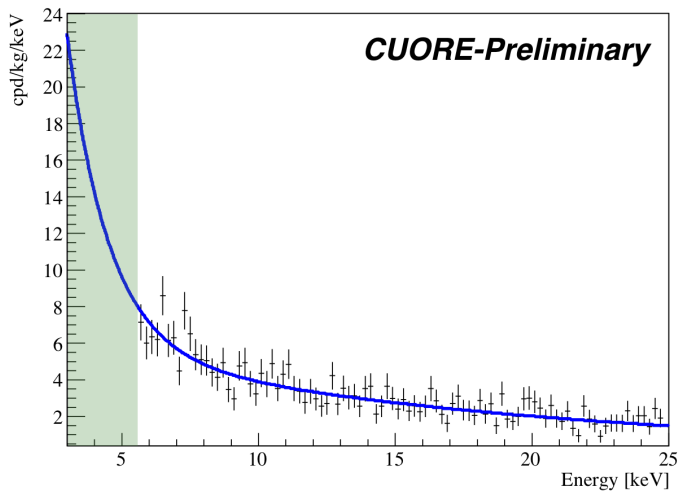}
\caption{Differential rate at low energies obtained combining data from the three bolometers having the energy threshold at 3\un{keV}.
The region below 5\un{keV} is not shown, since it is still under study (see text). The data are fitted with a double exponential decay, and the
fit function is used as model of the CUORE background.}
\label{fig:background}
\end{figure}

The four bolometers array took 19\un{days} of data. The low energy spectrum obtained combining the three bolometers having 3\un{keV} threshold, after cutting on the shape parameter and correcting for the detection efficiencies, is shown in  Fig.~\ref{fig:background}. 
The data in the range 3-5\un{keV} are not displayed, since they are still being studied; 
in place of the data we show their approximate behavior, that is an exponential decay. The counting rate ranges from $\sim 20\un{cpd/keV/kg}$ at $ 3\un{keV}$ to  $\sim 2\un{cpd/keV/kg}$ at $25\un{keV}$, and is well described by a double exponential decay. 
sWe estimate the \Cuore\ sensitivity to WIMPs assuming that the background differential rate will be equal
to the measured value in the four bolometers array, and that all bolometers will have a 3\un{keV} threshold. 
 It has to be stressed that \Cuore\ bolometers, unlike many other detectors, cannot distinguish nuclear recoils  from $\beta/\gamma$ interactions. 
Dark matter candidates like WIMPs, for example, are expected to induce nuclear recoils, so that experiments able to reject other
types of interactions have a higher discovery potential. Nevertheless \Cuore, thanks to its mass and to the low background
expected, could detect dark matter searching for an annual modulation of the coutning rate at low energies. 

\begin{figure}
\centering
\includegraphics[width=0.6\textwidth]{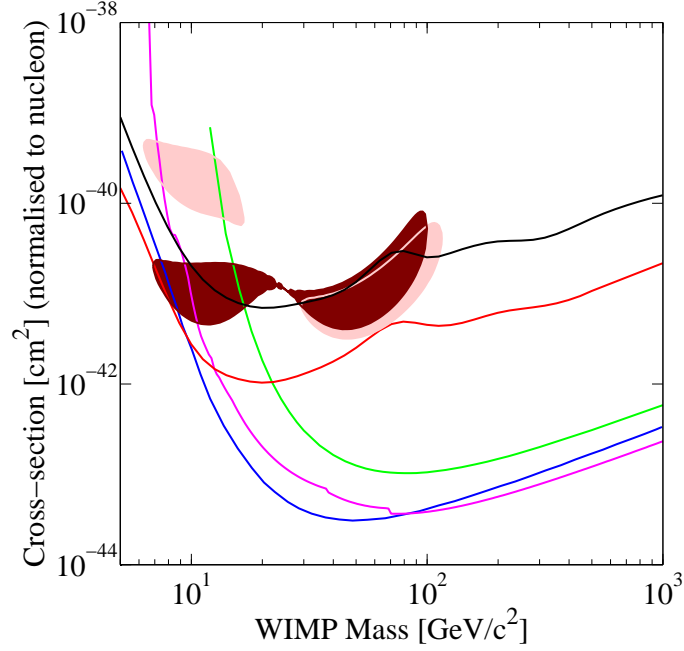}
\caption{CUORE(5y) (in red) and CUORE-0(3y) (in black) expected sensitivities compared to the actual 
limits on WIMP-nucleon elastic scattering cross section. DAMA/LIBRA 3$\sigma$ evidence with ion 
channeling in dark red and without ion channeling in pink, Edelweiss II in green, XENON100 
in blue, CDMS in magenta. 
}
\label{fig:exclusion}
\end{figure}

We performed toy Monte Carlo simulations generating background events from the fit of the measured distribution in Fig.~\ref{fig:background}, 
and WIMP events from the theoretical distribution described in Ref.~\cite{LewinSmith1996}, using a quenching factor for nuclear recoils in \TEO\ equal to 1~\cite{alessandrello:1998na}. We included the dependence of the WIMP interaction rate on the time in the year, and estimated the background+signal asymmetry subtracting the 3-months integrated spectrum across 2 December from the  3-months integrated spectrum across 2 June.
The 90\%\un{C.L.} sensitivity to the cross-section for spin independent interactions normalized to nucleon, as a function of the WIMP mass, for CUORE-0 in 3 years and CUORE in 5 years  of data taking,  is shown in Fig.~\ref{fig:exclusion} together with the present results of the leading experiments of the field. As it can be seen from the figure, \Cuore-0 and \Cuore\ will not be as sensitive as experiments able to discriminate nuclear recoils. \Cuore, however, could investigate the same parameter space of the DAMA/LIBRA experiment, and could be the only experiment other than DAMA/LIBRA looking for an annual modulation of dark matter interactions.

%\begin{thebibliography}{99}
\bibliographystyle{JHEP} % BibT
\bibliography{vignati} % BibTeX database without .bib extension 

%\end{thebibliography}

\end{document}